%Paper: hep-th/9205105
%From: schiff@guinness.ias.edu (Jeremy Schiff)
%Date: Thu, 28 May 92 09:49:18 EDT
%Date (revised): Tue, 6 Oct 92 14:14:25 EDT

%plain tex
\magnification=\magstep1
\baselineskip=18pt
\overfullrule=0pt
\nopagenumbers
\font\twelvebf=cmbx12

\rightline{hep-th/9205105}
\rightline{IASSNS-HEP-92/28}
\rightline{May 1992, revised Aug 1992}
\vskip .7in
\centerline{\twelvebf  The KdV Action and}
\centerline{\twelvebf Deformed Minimal Models}
\vskip 1in
\centerline{ Jeremy Schiff}
\smallskip
\centerline{\it Institute For Advanced Study}
\centerline{\it Olden Lane, Princeton, NJ 08540}
\vskip 1in
\centerline{\bf Abstract}
\smallskip
\noindent
An action is constructed that gives an arbitrary equation in the KdV or MKdV
hierarchies as equation of motion; the second Hamiltonian structure of
the KdV equation and the Hamiltonian structure of the MKdV equation appear as
Poisson bracket structures derived from this action. Quantization of this
theory can be carried out in two different schemes, to obtain either the
quantum KdV theory of Kupershmidt and Mathieu or the quantum MKdV theory of
Sasaki and Yamanaka. The latter is, for specific values of the coupling
constant, related to a generalized deformation of the
minimal models, and clarifies the relationship
of integrable systems of KdV type and conformal field theories. As a
generalization it is shown how to construct an action for the $SL(3)$-KdV
(Boussinesq) hierarchy.
\vfill\eject

\footline={\hss\tenrm\folio\hss}
\pageno=1

\noindent
An action for the KdV equation should have two basic properties:

\noindent
\item{(a)} The associated equation of motion should be the KdV
equation (or some equation in the KdV hierarchy).
\item{(b)} The associated Poisson bracket structure (we will
reiterate below how to derive Poisson brackets from an action) should
be the second hamiltonian structure of the KdV equation.

\noindent
We should be able to define a quantum theory using our action; given
(b) we might expect this to coincide in some sense with the
quantum KdV theory as described in [1]. This is clearly desirable,
given the correspondence of the conserved quantities of the
quantum KdV equation of [1] and the conserved quantities in deformed
minimal conformal theories. So we  add one further non-essential but
desirable property to our list above:

\noindent
\item{(c)} The Heisenberg equation of motion associated with our action
should be the quantum KdV equation of [1].

\noindent
In this note I construct an action that has properties
(a),(b),(c). The action can also be regarded as an action for the
MKdV equation. In this form the action has a kinetic term that describes
a theory which is ``nearly'' free, and an infinite number of potential
terms. In quantizing the theory defined by just the kinetic term, we
find many of the features of the Feigin-Fuchs construction for the
minimal models; in particular it becomes clear that the quantum analogs
of the terms in the potential (the quantum MKdV hamiltonians)
describe an infinite number of possible integrable deformations of
minimal conformal models.
These deformations are more general than those considered by Zamolodchikov [2].
Zamolodchikov's deformations of a conformal field theory
are ones that preserve both
the integrability and Lorentz invariance of the theory, and there are
an infinite number of other perturbations that  preserve just the
integrability. In the simplest case, the one that we shall consider,
the Zamolodchikov deformation gives rise to the integrable, Lorentz-invariant
Sine-Gordon theory (as recognized in [3]), whereas the deformations we
will consider give rise to theories with equations of motion in the
MKdV hierarchy (as is well-known, all the MKdV flows commute with the
Sine-Gordon flow). The correspondence of the conserved quantities of the
quantum KdV equation and the conserved quantities of deformed minimal
models becomes very clear.

In the last part of this note I also show how to construct
an action for the SL(3)-KdV equation.

Doubtless many physicists, on meeting the KdV equation for the first
time, investigate whether it can be derived from an action. It certainly
seems that the action has to be non-local in the KdV field $u$, and the
simplest actions one might guess appear unenlightening (see for example [4]).
Indeed, to satsify condition (b) above we need our action to be non-local
in $u$. An explanation of this is as follows: in a classical mechanical
system with phase space coordinates $X^i$, $i=1,...,2n$, a
hamiltonian structure is specified either by giving a non-degenerate
symplectic form on the phase space
$$ \Omega={\textstyle{1\over 2}}\omega_{ij} dX^i \wedge dX^j  \eqno{(1)}$$
or the corresponding set of Poisson brackets
$$ \{X^i,X^j\}~=~(\omega^{-1})^{ij} \eqno{(2)}$$
The second hamiltonian structure of the KdV equation
is given by the Poisson brackets
$$ \{u(x),u(y)\}~=~-{\textstyle{{24\pi}\over{c}}}
               \bigl(\partial_x^3+u(x)\partial_x+\partial_xu(x)
               \bigr)\delta(x-y)     \eqno{(3)}$$
and we see at once that the corresponding symplectic form must
contain a highly non-local operator. But if the action is local, then
the symplectic form must also be local.

The above observations also suggest a way to look for a good action.
Wilson [5] has noted that while the symplectic form associated with the
second hamiltonian structure of the KdV equation is non-local, the
symplectic form associated with the corresponding hamiltonian structure
of the ``Ur-KdV equation'' (the name is Wilson's) {\it is} local.
I review his result. Any solution $q$ of the Ur-KdV equation
$$ q_t~=~q_{xxx} - {\textstyle{3\over 2}}q_{xx}^2q_x^{-1}  \eqno{(4)}$$
gives a solution of the KdV equation
$$ u_t~=~u_{xxx} + 3 u u_x                                 \eqno{(5)}$$
via the ``Miura map''
$$ u~=~\{q;x\}~=~q_{xxx}q_x^{-1}-{\textstyle{3\over 2}}q_{xx}^2q_x^{-2}
                             \eqno{(6)}$$
($\{q;x\}$ denotes the Schwarzian derivative of $q$ with respect to $x$).
If we take the brackets
$$ \{q(x),q(y)\}~=~{\textstyle{{24\pi}\over{c}}}
                  \partial_x^{-1}q_x\partial_x^{-1}q_x\partial_x^{-1}
                  \delta(x-y)   \eqno{(7)}$$
then $u$ defined by (6) will satisfy (3). Now the {\it inverse} of the
operator preceding the delta function on the right hand side of (7) is
clearly local, i.e. the associated symplectic form is local.
So we look for an action that is local in the function $q$.

One action that gives the correct symplectic form/Poisson brackets
is the geometric Virasoro action of Polyakov,
Bershadsky, Ooguri and others [6]:
$$ S_0=-{c \over {48\pi}}\int dxdt~~q_{xt}q_{xx}q_x^{-2}
                                 \eqno{(8)}$$
At this point I briefly digress to give an account (which I learned from
V.P.Nair) of how to find the symplectic form determined by an action. Suppose,
for definiteness, that we have an action $S$ for a single field $\phi$
in $1+1$ dimensions. Integrating by parts if necessary, we can find
an expression for the variation of the action in the form
$$ \delta S = \int dxdt ~~\biggr(p[\phi] \delta\phi + \partial_t\alpha
              + \partial_x\beta \biggr) \eqno{(9)}$$
Here $p[\phi]$ is some density depending on $\phi$ and its derivatives,
and $\alpha$ and $\beta$ are densities which depend on $\phi$ and its
derivatives, and also are linear in the variation of $\phi$ and its
derivatives. The first term in (9) yields the equation of motion $p=0$,
but there is clearly further information. From the term in (9) which is
a total derivative with respect to the time $t$, we obtain a one-form
on the space of functions $\phi$ which are independent of $t$:
$$ \tilde{\alpha}~=~\int dx~~\alpha \eqno{(10)}$$
Adding a total derivative term to the action $S$ would change
$\tilde{\alpha}$ (it does not, of course, change the equations of motion),
but it is easy to see that the term that would be added would be exact,
so the two-form
$$ \Omega~=~\delta\tilde{\alpha}  \eqno{(11)}$$
is unaffected. This is the symplectic form determined by $S$ (or, more
precisely, determined by $S$ and a choice of ``time'' direction).
Using this method it is easy to check that the action (8) gives the
symplectic form associated with the brackets
(7). Now, classically the action $S_0$  just describes
a free theory, since writing $h=\ln q_x$ we have
$$ S_0 = -{c \over {48\pi}}\int dxdt~~h_x h_t \eqno{(12)}$$
So $S_0$ is clearly not a candidate KdV action. But we can add to $S_0$
terms dependent only on $q$ and its $x$-derivatives {\it without}
changing the Poisson brackets. Before we do this, though, we
note the other crucial property of $S_0$, that (ignoring the boundary
terms crucial for the derivation of the symplectic form)
$$ \delta S_0 = - {c \over {24\pi}}
               \int dxdt~~u_t q_x^{-1} \delta q \eqno{(13)}$$
Thus the equation of motion derived from an action with ``kinetic''
term $S_0$ will give evolution equations for $u$, just as we desire.

To complete the construction is now easy. Let  $p[u]$ be some
density in $u$ and its derivatives; write
$$ H~=~\int dx~ p[u] \eqno{(14)}$$
and define $\delta p/\delta u$ by
$$ \delta H~=~\int dx~{{\delta p}\over{\delta u}}\delta u \eqno{(15)}$$
(Throughout this work we take $x$ to be defined on some finite range,
and assume all functions to satisfy periodic boundary conditions, so that we
can integrate by parts with respect to $x$ without boundary terms appearing).
Then we find
$$ \delta H~=~-\int dx~~ {{\delta q}\over{q_x}}
            \bigl(\partial_x^3+u(x)\partial_x+\partial_xu(x)\bigr)
            {{\delta p}\over{\delta u}}
                   \eqno{(16)}$$
So we consider the action
$$ S~=~S_0~+~\sum_{n=1}^{\infty}\lambda_n\int dxdt~p_n[u]  \eqno{(17)}$$
where the $\lambda_n$, $n=1,2,...$,  are constants and the $p_n[u]$,
$n=1,2,...$, are the densities of the conserved quantities of the KdV
equation (see, for example, [4]) e.g.
$$ \eqalign{ p_1[u] &= u \cr
             p_2[u] &= {\textstyle {1 \over 2}} u^2 \cr
             p_3[u] &= {\textstyle {1 \over 2}} (u^3-u_x^2) \cr}
                    \eqno{(18)}$$
The $p_n[u]$ are related by the Lenard recursion relation
$$ \partial_x {{\delta p_n}\over{\delta u}} =
     \bigl(\partial_x^3+u(x)\partial_x+\partial_xu(x)\bigr)
     {{\delta p_{n-1}}\over{\delta u}} \eqno{(19)}$$
$S$ is the classical KdV action, which has properties (a)
and (b) we listed at the start of this paper; indeed any equation in the
KdV hierarchy (or any ``linear combination'' of the equations, in the
obvious sense) can be obtained by suitable choice of the constants
$\lambda_n$.

To quantize the theory defined by $S$ requires a little care, but is
essentially straightforward thanks to existing results in the literature.
We start by discussing the action $S_0$. To quantize any theory, we
select a set of local Poisson brackets and elevate it to the level of
an operator commutation relation. As we will see below (3) is not the
only set of local Poisson brackets we can derive from $S_0$. But to
obtain the quantum KdV theory of [1] we indeed choose (3) as our ``fundamental
bracket'', which now becomes an operator commutation relation.
Comparing (3) with the formulae of Gervais [7],
we see our quantum theory is characterized by a Virasoro algebra of
central charge $c$. To obtain $S$ from $S_0$ in the classical theory
we added a ``potential'' consisting of an infinite sum of terms
proportional to the  conserved quantities of the
classical KdV equation. Already in [7] Gervais conjectured that
there are quantum analogs of these, i.e. given that the {\it operator}
field $u(x)$ satisfies the commutation relations associated with the
bracket (3), for an arbitrary central charge $c$, there exist an infinite
number of operators ${\cal P}_n$, $n=1,2,...$, all integrals of
normal-ordered densities in $u$ and its derivatives, that mutually commute.
Gervais' conjecture received
substantial support from the work of Sasaki and Yamanaka [8], who
computed ${\cal P}_1,{\cal P}_2,{\cal P}_3,{\cal P}_4,{\cal P}_5$, and it
was finally proved in [9] (see also [10]). Thus the obvious way to define
quantum KdV theory is, as in [1], via the Heisenberg equation of motion
$$ u_t=\left[u,\sum_{n=1}^{\infty}\lambda_n{\cal P}_n\right] \eqno{(20)}$$
The equation of [1] is just this with $\lambda_2=1$, and $\lambda_n=0$
for $n\not=2$.

At this juncture it is probably appropriate to point out that the
title ``the conserved quantities of the quantum KdV equation''
for the operators ${\cal P}_n$ is somewhat
misleading. The operators ${\cal P}_n$ exist because
the modes of $u(x)$ satisfy a Virasoro algebra. Quantum KdV theory
can only be defined {\it because} the mutually commuting
operators ${\cal P}_n$ exist, and
is defined in a manner that makes it obvious that the ${\cal P}_n$'s
remain conserved quantities. Calling them
``the conserved quantities of the quantum KdV equation'' is therefore
putting the cart before the horse.

Returning to the classical theory, our next observation is that
the action $S$ can also be considered as an action for the
field $h=\ln q_x$. $S_0$ is given in (12), and for the other terms we
just use $u=h_{xx}-{1 \over 2}h_x^2$. Introducing $j=h_x$, the equation
of motion is
$$ {c \over {24\pi}}j_t=-\partial_x(\partial_x+j)\sum_{n=1}^{\infty}
           \lambda_n{{\delta p_n}\over{\delta u}}  \eqno{(21)}$$
This is the general equation in the MKdV hierarchy.
The Poisson bracket for $j$ is simply
$$ \{j(x),j(y)\}~=~{{24\pi}\over c}
                   \partial_x\delta(x-y) \eqno{(22)}$$
which is the usual Poisson bracket structure for the MKdV hierarchy.
The equation of  motion (21) for $j$ implies the equation of motion for
$u=j_x-{1 \over 2}j^2$, but of course the equation for $u$ does not
imply that for $j$, since varying $h$ is more general than varying $q$.
We will from now on mostly be interested in $S$ as an action for
$h$, but before we proceed we note that we can also regard $S$ as a
{\it non-local} action for either $j$ or $u$. We find
$$ \delta S_0~=~-{c \over {24\pi}}\int dxdt~h_t\delta j
             ~=~{c \over {24\pi}} \int dxdt~q_tq_x^{-1}\delta u \eqno{(23)}$$
so regarding $S$ as an action for $j$ yields the equation of motion
$$ {c \over {24\pi}}h_t=-(\partial_x+h_x)\sum_{n=1}^{\infty}\lambda_n
                      {{\delta p_n}\over{\delta u}}  \eqno{(24)}$$
and regarding it as an action for  $u$ yields the equation of motion
$$ {c \over {24\pi}}q_t=-q_x\sum_{n=1}^{\infty}\lambda_n
                      {{\delta p_n}\over{\delta u}}  \eqno{(25)}$$
This last equation is the general equation in the Ur-KdV hierarchy. The
hierarchy defined in (23) is known as the potential KdV hierarchy.

Now, when we regard $S$ as an action for the field $h$ it turns out to be
natural to add two further terms to the action.
There are two more local functionals of $h$ which
commute with all the classical KdV hamiltonians $\int dxdt~p_n$, viz.
$V_+=\int dxdt~e^h$ and $V_-=\int dxdt~e^{-h}$ (to prove
that $V_-$ commutes with all the KdV hamiltonians requires use of the
result that the KdV hamiltonians are symmetric under $j\rightarrow -j$).
So we consider the modified action
$$ S_M~=~S+\lambda_+V_++\lambda_-V_-  \eqno{(26)}$$
$S$ is invariant under a constant shift of the field $h$, so if
$\lambda_+$ and $\lambda_-$ are both non-zero we can without loss of
generality take them to be equal. If one of $\lambda_+$,$\lambda_-$ is
zero, then since $S$ is invariant under $h\rightarrow -h$ we can without
loss of generality take it to be $\lambda_-$ that is zero, and then by
shifting $h$ we can set $\lambda_+=1$.
Thus we see the effect of adding these
terms to the equation of motion (21); they add either a term proportional
to $\sinh h$ or a term $e^h$ on the right hand side of (21). Thus the
action $S_M$ is an action for the general MKdV/Sinh-Gordon flow or the
general MKdV/Liouville flow. Note that the quantities $V_+$ and $V_-$
{\it do not} commute with each other. Note also that we might have
considered adding terms proportional to $V_+$ and $V_-$ to $S$ considered
as an action for $q$; but $V_+$ vanishes when we set $h=\ln q_x$ and assume
periodic boundary conditions for $q$, and $V_-$ gives an extra evolution
equation for $u$ which can not be expressed simply in terms of $u$, but
requires use of the function $q$.

The first step in quantization of the theory defined by $S_M$ is to
quantize $S_0$ by using the Poisson bracket for the field $j$ as our
``fundamental bracket''. At first glance this is just quantization of
a free theory. But let us try to quantize the theory ``remembering'' that
$h=\ln q_x$. If we take to $q$ to satisfy periodic boundary conditions
(in $x$), and we also want $h$ to satisfy periodic boundary conditions,
this will have some important consequences. First, assuming we want
$S_0$ to be real, we need $h$ to be either real or pure imaginary; the first
possibility is not consistent with periodic boundary conditions on $q$,
so we must take $h$ pure imaginary (this implies a very nasty
constraint on $q$, but this need not concern us). Let us write $h=-i\beta\phi$
and assume $c$ is positive, so that by correct choice of $\beta$ we can take
$$ S_0~=~{1\over{8\pi}}\int dxdt~\phi_x\phi_t \eqno{(27)}$$
Next, having decided that $h$ is imaginary, we recall that the imaginary
part of a logarithm is only defined mod $2\pi$, so $\phi$ can only be
defined mod $2\pi/\beta$ (i.e. it is a compactified field). This is
in fact good. Supposing the range of $x$ to be $2\pi L$, it is clear that
we could take $q=(L/in)e^{inx/L}$, where $L$ is an integer. This would give
$\phi=-nx/L\beta$, which is not periodic unless $\phi$ is a compactified
field and $L$ is restricted\footnote*{This seems very strange; but it must be
remembered that in writing (27) we have fixed the coupling constant of the
theory, and we can in fact formulate the theory with an arbitrary range for
$x$ but with a specific coupling constant.}. The final deduction we
can make from the $h=\ln q_x$ relation is that $\int dx~e^{-i\beta\phi}=0$
must vanish. Thus we should investigate the quantum theory of a compactified
field satisfying free field commutation relations (determined from (27))
subject to the constraint $\int dx~e^{-i\beta\phi}=0$, where $\beta^{-1}$
is the radius of compactification.

I do not intend here to pursue this quantization to the end. The main
point we need is that states in the theory will be defined as states in
a free field theory that are in some sense
annihilated by the (normal ordered) constraint\footnote{**}{Looking at the
work of Felder [11] we see that that to make this precise it is
necessary to define a somewhat extended action of the constraint
operator. I do not intend to enter into the details of this here.}, and
permitted operators which are polynomial in $\phi$ and its $x$ derivatives
will have to commute with the constraint. In
particular, if we seek a permitted operator of form
$$ T=-{\textstyle{1\over 4}}\phi_x^2+i\alpha\phi_{xx} \eqno{(28)}$$
we find we need $\alpha={1\over 2}(\beta-\beta^{-1})$. The modes of
$T$ satisfy a Virasoro algebra of central charge
$\tilde{c}=13-6(\beta^2+1/\beta^2)$, (reproducing (4.18) in the second
paper of [8], with $\hbar=1/2$). For $\beta=\sqrt{m/(m+1)}$, $m=3,4,...$, this
reproduces the central charges of the unitary minimal models. We see thus
that the quantum theory based on $S_0$ constructed thus is very much
related to the Feigin-Fuchs construction for the minimal models.

It is straightforward now to exploit the results
of [8] to understand the full quantum theory associated
with $S_M$. First we note that since we have identified a Virasoro field
in the theory defined by $S_0$, we can add to it the operators ${\cal P}_n$
defined above appropriate for the central charge $\tilde{c}$.
This gives us the quantum version of $S$.
For the quantum analogs of $V_{\pm}$, which we will call ${\cal V}_{\pm}$,
we take the normal ordered versions of their classical
counterparts, ${\cal V}_{\pm}=\int dxdt~:e^{\mp i\beta\phi}:$.
Clearly the ${\cal P}_n$'s, which are constructed out of $T$,
all commute with ${\cal V}_+$ and since the ${\cal P}_n$ are symmetric under
$\phi\rightarrow -\phi$ (as in the classical case), they must therefore
also commute with ${\cal V}_-$\footnote{***}{Again, some deeper
analysis is required to understand in what sense ${\cal V}_-$ is defined
in the theory.}. The significance of the commutation of ${\cal V}_{-}$
with the ${\cal P}_n$'s in conformal field theory is exactly the statement
that the quantum MKdV hamiltonians are conserved quantities in the
$\Phi_{(1,3)}$ perturbation of the minimal models [2]; this is because
(as noted by Eguchi and Yang [3]) in the Feigin-Fuchs formulation of the
minimal models with energy momentum tensor $T$,
the vertex operator  $:e^{i\beta\phi}:$ is just the $(1,3)$ primary field
(of conformal dimension $(m-1)/(m+1)$) ($:e^{-i\beta\phi}:$ is one of
the screening currents, of dimension $1$). We see in fact that the most
general integrable perturbation of the minimal models including the
$\Phi_{(1,3)}$ perturbation consists of adding terms proportional
to ${\cal P}_n$ and  ${\cal V}_+$.
While it is a somewhat obvious statement that you can add to the hamiltonian
of an integrable field theory an arbitrary sum of the conserved quantities,
the fact that quantum MKdV theory is an integrable perturbation of
the minimal models seems not to be widely appreciated.

\vskip.2in

\noindent{\bf Generalizations}

\noindent
We would naturally like to extend this work to the $SL(N)$-KdV equations
for $N>2$, and ultimately to other integrable systems, both for the sake
of having actions for integrable systems {\it per se} and for the sake of
understanding all integrable deformations of conformal field theories.
Here I only intend to briefly present the construction of the action for
the $SL(3)$-KdV case, i.e. I discuss actions for the Boussinesq hierarchy; it
seems likely that this will extend smoothly to the $SL(N)$ case.

The crucial ingredients are the analog of $S_0$ (which we will call
$S_0^{(3)}$) and the identification
of appropriate fields, the analogs of $q,h,j,u$ from above. We expect
$S_0^{(3)}$ to be a constrained WZW action
of the kind discussed in [6], which in appropriate variables
(the analogs of $h$ from above) is a free
field theory. Explicitly, we take
$$ S_0^{(3)}~=~\nu\int dxdt~\left(\sum_{i=1}^3 h_{ix}h_{it}\right)
       \eqno{(28)}$$
where $\nu$ is a constant and $h_1,h_2,h_3$ are three fields
satisfying $h_1+h_2+h_3=0$.
If we eliminate $h_2$ from $S_0^{(3)}$ it
essentially reproduces equation (166) in [6]. We define $j_i=\partial_xh_i$,
$i=1,2,3$. The analogs of the field $u$ from above will presumably be
related to the fields $j_i$ by a standard Miura map. The only question that
needs to be resolved is the identification of the analogs of $q$.

The analogs of $q$ should provide us with a parametrization of
$SL(3)$ matrices $g$ such that
$$ [g^{-1}g_x]_+~=~\pmatrix{0&1&0\cr
                            0&0&1\cr
                            0&0&0\cr} \eqno{(29)}$$
Here $[M]_+$ denotes the strictly upper triangular part of $M$.
This subset of $SL(3)$ is invariant under $g\rightarrow gU$, where
$$ U~=~\pmatrix{1&0&0\cr
                \alpha&1&0\cr
                \beta&\gamma&1\cr} \eqno{(30)}$$
Taking inspiration from the further work of Wilson [12], we observe
that if (29) is satisfied we can write
$$ g~=~\pmatrix{a_{xx}+Aa_x+Ba & a_x+Ca & a \cr
                b_{xx}+Ab_x+Bb & b_x+Cb & b \cr
                c_{xx}+Ac_x+Bc & c_x+Cc & c \cr} \eqno{(31)}$$
where $a,b,c,A,B,C$ are functions, with the Wronskian of $a,b,c$ equal to $1$.
The group action on $g$ essentially corresponds to translations of $A,B,C$.
To complete the parametrization we just need a parametrization of the three
functions $a,b,c$ whose Wronskian is $1$. One way to obtain this is to write
$$\eqalign{a&=c\phi\cr
           b&=c\psi}    \eqno{(32)}$$
Then the Wronskian condition is solved to give
$ c=(\phi_{xx}\psi_x-\psi_{xx}\phi_x)^{-1/3} $
and thus $a,b,c$ are written in terms of $\phi,\psi$.

We call different choices of $A,B,C$ in $g$ different ``gauges''. Two gauges
are of particular interest; if $A,B,C$ are set to zero we find
$$ g^{-1}g_x~=~\pmatrix{0&1&0\cr
                        u_2&0&1\cr
                        u_3&0&0\cr} \eqno{(33)}$$
and if $A,B,C$ are chosen so that $g$ is upper triangular (which is clearly
possible if $bc_x-cb_x\not=0$)
$$ g^{-1}g_x~=~\pmatrix{j_1&1&0\cr
                        0&j_2&1\cr
                        0&0&j_3\cr} \eqno{(34)}$$
with $j_1+j_2+j_3=0$. These formulae complete the relationships
between the different variables we need, which I now summarize:
$$\eqalign{h_1&=-\ln(c^2\psi_x)\cr
           h_3&=\ln c\cr}\eqno{(35)}$$
$$ c=(\phi_{xx}\psi_x-\psi_{xx}\phi_x)^{-{\textstyle{1\over 3}}}
                    \eqno{(36)}$$
$$j_i=\partial_xh_i,~~~i=1,3\eqno{(37)}$$
$$\eqalign{u_2&=(j_3-j_1)_x+j_1^2+j_1j_3+j_3^2\cr
           u_3&=j_{3xx}+j_3(2j_3+j_1)_x-j_1j_3(j_1+j_3)\cr} \eqno{(38)}$$

All that remains is to do some hard calculations.
We find (ignoring boundary contributions)
$$ \eqalign{\delta S_0^{(3)}
&=-2\nu\int dxdt~~(u_{3t}-u_{2xt}+j_1u_{2t})\psi_xc^3\delta\phi
       - (u_{3t}-u_{2xt}+Lu_{2t})\phi_xc^3\delta\psi\cr
&=-2\nu\int dxdt~~(2j_1+j_3)_t\delta h_1 + (2j_3+j_1)_t\delta h_3\cr
&=~2\nu\int dxdt~~(2h_1+h_3)_t\delta j_1 + (2h_3+h_1)_t\delta j_3\cr
&=~2\nu\int dxdt~~(\phi_t\psi_x-\psi_t\phi_x)c^3\delta u_3 +
  (\psi_x(c\phi_t)_x-\phi_x(c\psi_t)_x)c^2\delta u_2  }\eqno{(39)}$$
where $L=-\partial_x\ln(c^2\phi_x)$.
The local Poisson brackets derived from $S_0^{(3)}$ are
$$ \pmatrix{\{j_1(x),j_1(y)\}&\{j_1(x),j_3(y)\}\cr
            \{j_3(x),j_1(y)\}&\{j_3(x),j_3(y)\}\cr}
   = -{1 \over {6\nu}}\pmatrix{2&-1\cr
                               -1&2\cr}\partial_x\delta(x-y) \eqno{(40)}$$
$$ \displaylines{
   \pmatrix{\{u_2(x),u_2(y)\}&\{u_2(x),v_3(y)\}\cr
            \{v_3(x),u_2(y)\}&\{v_3(x),v_3(y)\}\cr} =\hfill\cr
   \hfill{} -{1 \over {2\nu}}\pmatrix{
    -2\partial_x^3+u_2\partial_x+\partial_xu_2&
    v_3\partial_x+2\partial_xv_3\cr
    2v_3\partial_x+\partial_xv_3&
    {{\textstyle{2\over 3}}\partial_x^5
     -{\textstyle{5\over 3}}(u_2\partial_x^3+\partial_x^3u_2)
     +(u_{2xx}\partial_x+\partial_xu_{2xx})
     +{\textstyle{8\over 3}}u_2\partial_xu_2}\cr}\delta(x-y) \cr}$$
$$\eqno{(41)}$$
In (41) I have made a standard field redefinition, setting
$v_3=2u_3-u_{2x}$. The matrix in (41) agrees, up to a trivial rescaling,
with the second hamiltonian structure of the Boussinesq equation as given
in [13], example 7.28. From (40) we see that the combinations
$J=j_1+j_3$ and $K=j_1-j_3$ commute with respect to the Poisson bracket,
and these are useful for certain calculations. From the $1,1$ entry of
(41) and the formulae of [7] we identify the central charge that will
charcterize quantum Boussinesq theory to be $-96\pi\nu$ (this calculation
actually verifies a conjecture made in [6b] just before equation (173);
our action is related to the action of [6b] by $S_0^{(3)}=4\pi\nu S_{W_3}$).

The full Boussinesq action is taken to be
$$ S^{(3)}~=~S_0^{(3)}~+~\sum_{{n=1}\atop{n\not\equiv 0~mod~3}}^{\infty}
      \lambda_n\int dxdt~p_n[u_2,v_3] \eqno{(42)}$$
where $p_n[u_2,v_3]$, $n=1,2,4,5,...$, are the densities of the conserved
quantities of the Boussinesq equation. These satisfy
$$ \displaylines{
  \pmatrix{0&\partial_x\cr\partial_x&0\cr}
  \pmatrix{{{\delta p_{n+3}}\over{\delta u_2}}\cr
           {{\delta p_{n+3}}\over{\delta v_3}}\cr} = \hfill\cr
  \hfill{} \pmatrix{
    -2\partial_x^3+u_2\partial_x+\partial_xu_2&
    v_3\partial_x+2\partial_xv_3\cr
    2v_3\partial_x+\partial_xv_3&
    {{\textstyle{2\over 3}}\partial_x^5
     -{\textstyle{5\over 3}}(u_2\partial_x^3+\partial_x^3u_2)
     +(u_{2xx}\partial_x+\partial_xu_{2xx})
     +{\textstyle{8\over 3}}u_2\partial_xu_2}\cr}
        \pmatrix{{{\delta p_n}\over{\delta u_2}}\cr
           {{\delta p_n}\over{\delta v_3}}\cr} \cr}$$
$$\eqno{(43)}$$
The first few $p_n$'s are
$$ \eqalign{ p_1~&=~u_2\cr
             p_2~&=~v_3\cr
             p_4~&=~u_2v_3\cr
             p_5~&=~v_3^2+{4 \over 9}u_2^3-{1\over 3}u_{2x}^2} \eqno{(44)}$$
All the equations of the Boussinesq, modified Boussinesq and Ur-Boussinesq
hierarchies are found as the equations of motion, treating the action as
a functional of appropriate variables. I just illustrate for the case where
$\lambda_2=\nu$ and all the other $\lambda_n$ are zero; we obtain the
equations
$$ \eqalign{u_{2t}&=-v_{3x}\cr
            v_{3t}&={\textstyle{1\over 3}}u_{2xxx}-
                     {\textstyle{2\over 3}}(u_2^2)_x   }\eqno{(45)}$$
$$ \eqalign{j_{1t}&={\textstyle{1\over 3}}
             (2j_{3x}+j_{1x}+2j_3^2-j_1^2+2j_1j_3)_x\cr
            j_{3t}&={\textstyle{1\over 3}}
             (-2j_{1x}-j_{3x}+2j_1^2-j_3^2+2j_1j_3)_x  }  \eqno{(46)}$$
$$ \eqalign{\phi_t&=-\phi_{xx}-2c^{-1}c_x\phi_x\cr
            \psi_t&=-\psi_{xx}-2c^{-1}c_x\psi_x  } \eqno{(47)}$$
where in (47) $c$ is given by (36). Equation (46) is transformed into
equation (2.7) of [14] via the substitution
$$ \pmatrix{p_1\cr p_2}=\pmatrix{3^{-1/2}&-3^{-1/2}\cr 1&1}
                        \pmatrix{j_1\cr j_3}
                 \eqno{(48)}$$
and by rescaling the coordinates.
It remains to remark that we can add to the action $S^{(3)}$ a sum of terms
proportional to $\int dxdt~\exp(2h_1+h_3)$, $\int dxdt~\exp(-h_1-2h_3)$
and $\int dxdt~\exp(h_3-h_1)$, without ruining the integrability. These
terms give rise to the $SL(3)$-Toda flow that commutes with the MKdV flows;
note the first two terms vanish upon writing $h_1$ and $h_3$ in terms of
$\phi$ and $\psi$. This completes construction of the classical $SL(3)$-KdV
action.

\vskip.2in

\noindent{\bf Conclusions}

\noindent The work presented here goes some way towards clarifying the
magical properties of both classical and quantum integrable systems.
We started our constructions with the choice of a ``free'' action ($S_0$ or
$S_0^{(3)}$) which, being first order in time derivatives, determined a
Poisson bracket algebra but gave a vanishing Hamiltonian. This Poisson bracket
structure (and the corresponding set of operator commutation relations) has
the property that it admits an infinite number of mutually commuting
quantities of the type now familiar. Of course, this is not a simple result
(particularly in the quantum case), and I have not here said anything about the
proof of this result. But it seems important to stress that once this property
has been established, as a property of the Poisson brackets/commutation
relations of the ``free'' theory, the existence and integrability of classical
and quantum KdV hierarchies becomes essentially obvious; it is just necessary
to add a Hamiltonian consisting of a sum of the mutually commuting quantities.

As I have already pointed out, it is of interest to generalize this work
to find actions for other integrable systems. One essential part of this is to
examine if and how Wilson's antiplectic formalism can be extended to systems
such as the non-linear Schr\"odinger hierarchy, and this will be tackled in a
forthcoming paper. Another question outstanding is to ask, since we have
obtained
quantum MKdV theory as a deformation of the minimal models, whether there is
a statistical mechanical meaning to such deformations. If so, it may be
interesting to see if via conformal field theoretical techniques one can do
computations in quantum MKdV theory.

One aspect of KdV theory that has been completely absent from this paper
is the $\tau$-function formalism. The relationship of the KdV field $u$
and the $\tau$-function is $u=2\partial_x^2\ln\tau$, but there is no simple
relation between $\tau$ and any of the fields $q,h,j$, so it is not clear
how to write an action to yield Hirota's form of the KdV equation (note though
that if we write $q_x=p^2$ we find $u=2p\partial_xp^{-1}\partial_x\ln p$).
The relationship between integrability of the KdV hierarchy from the point of
view of the existence of an infinite number of conserved quantities, and
integrability from the point of view of an infinite-dimensional group action
on the space of solutions remains somewhat mysterious.

\vskip.2in

\noindent{\bf Acknowledgements}

\noindent
I thank D.Depireux for bringing my attention to references [1],[5],[10] and
[12], and for some comments.
Useful discussions with R.Dijkgraaf, A.Giveon, O.Lechtenfeld,
V.P.Nair, E.Verlinde
and E.Witten are acknowledged. This work was supported by the U.S.Department
of Energy under grant DE-FG02-90ER40542.

\vskip.2in

\noindent{\bf References}

\noindent
\item{[1]} B.A.Kupershmidt and P.Mathieu, {\it Phys.Lett.B} {\bf 227}
              (1989) 245.
\item{[2]} A.B.Zamolodchikov, {\it Adv.Stud.in Pure Math.} {\bf 19}
      (1989) 641.
\item{[3]} T.Eguchi and S.Yang {\it Phys.Lett.B} {\bf 224} (1989) 373;
      T.Hollowood and P.Mansfield, {\it Phys.Lett.B} {\bf 226} (1989) 73.
\item{[4]} A.Das, {\it Integrable Models}, World Scientific (1989).
\item{[5]} G.Wilson, {\it Phys.Lett.A} {\bf 132} (1988) 45.
\item{[6]} a) A.M.Polyakov, {\it Mod.Phys.Lett.A} {\bf 2} (1987) 893;
        b) M.Bershadsky and H.Ooguri, {\it Comm.Math.Phys.} {\bf 126}
       (1989) 49 and references therein.
\item{[7]} J.-L. Gervais, {\it Phys.Lett.B} {\bf 160} (1985) 277.
\item{[8]} R.Sasaki and I.Yamanaka, {\it Comm.Math.Phys.} {\bf 108}
    (1987) 691, {\it Adv.Stud.in Pure Math.} {\bf 16} (1988) 271.
\item{[9]} B.Feigin and E.Frenkel, {\it Phys.Lett.B} {\bf 276} (1992) 79.
\item{[10]} P.Di Francesco, P.Mathieu and D.S\'en\'echal,
     {\it Mod.Phys.Lett.A} {\bf 7} (1992) 701.
\item{[11]} G.Felder, {\it Nucl.Phys.B} {\bf 317} (1989) 215.
\item{[12]} G.Wilson, {\it Quart.J.Math.Oxford} {\bf 42} (1991) 227,
      {\it Nonlinearity} {\bf 5} (1992) 109, and in {\it Hamiltonian
     Systems, Transformation Groups and Spectral Transform Methods},
     ed. J.Harnad and J.E.Marsden, CRM (1990).
\item{[13]} P.J.Olver {\it Applications of Lie Groups to Differential
    Equations}, Springer-Verlag (1986).
\item{[14]} P.Mathieu and W.Oevel, {\it Mod.Phys.Lett.A} {\bf 6} (1991) 2397.

\bye